\documentclass[aps,pra,10pt,a4paper,twocolumn,footinbib, superscriptaddress, floatfix]{revtex4-2}
\usepackage{graphicx}
\usepackage{amsmath}
\usepackage{amsfonts}
\usepackage{amssymb}
\usepackage{soul}
\usepackage{dsfont}
\usepackage{hyperref}
\usepackage{amstext}
\usepackage[caption=false]{subfig}
\usepackage{mathtools}
\usepackage{balance}
\hypersetup{colorlinks=true, citecolor=blue, urlcolor=blue, linkcolor=blue}
\usepackage{color,soul}
\usepackage{xurl}
\usepackage{physics}
\usepackage{braket}
\usepackage{siunitx}
\usepackage{floatrow}
\usepackage{xr}

\date{October 29, 2024}

\begin{document}

\title{Sensing force gradients with cavity optomechanics while evading backaction}

\author{Elisabet K. Arvidsson}
\affiliation{Department of Applied Physics, KTH Royal Institute of Technology, Hannes Alfvéns väg 12, SE-114 19 Stockholm, Sweden}
\author{Ermes Scarano}
\affiliation{Department of Applied Physics, KTH Royal Institute of Technology, Hannes Alfvéns väg 12, SE-114 19 Stockholm, Sweden}
\author{August K. Roos}
\affiliation{Department of Applied Physics, KTH Royal Institute of Technology, Hannes Alfvéns väg 12, SE-114 19 Stockholm, Sweden}
\author{Sofia Qvarfort}
\affiliation{Nordita, KTH Royal Institute of Technology and Stockholm University, Hannes Alfvéns väg 12, SE-114 19 Stockholm, Sweden}
\affiliation{Department of Physics, Stockholm University, AlbaNova University Center, SE-106 91 Stockholm, Sweden}
\author{David B. Haviland}
\email{haviland@kth.se}
\affiliation{Department of Applied Physics, KTH Royal Institute of Technology, Hannes Alfvéns väg 12, SE-114 19 Stockholm, Sweden}

\begin{abstract}

We study force-gradient sensing with a coherently driven mechanical resonator and phase-sensitive detection of motion through the two-tone backaction evading measurement of cavity optomechanics. The response of the optomechanical system, solved by numerical integration of the classical equations of motion, shows an extended region which is monotonic to changes in force gradient. We use Floquet theory to model the fluctuations, which rise only slightly above that of the usual backaction evading measurement in the presence of the mechanical drive. The monotonic response and minimal backaction are advantageous for applications such as atomic force microscopy.

\end{abstract}

\maketitle

\section{Introduction}

\begin{figure}[t]
\includegraphics[width=0.8\textwidth]{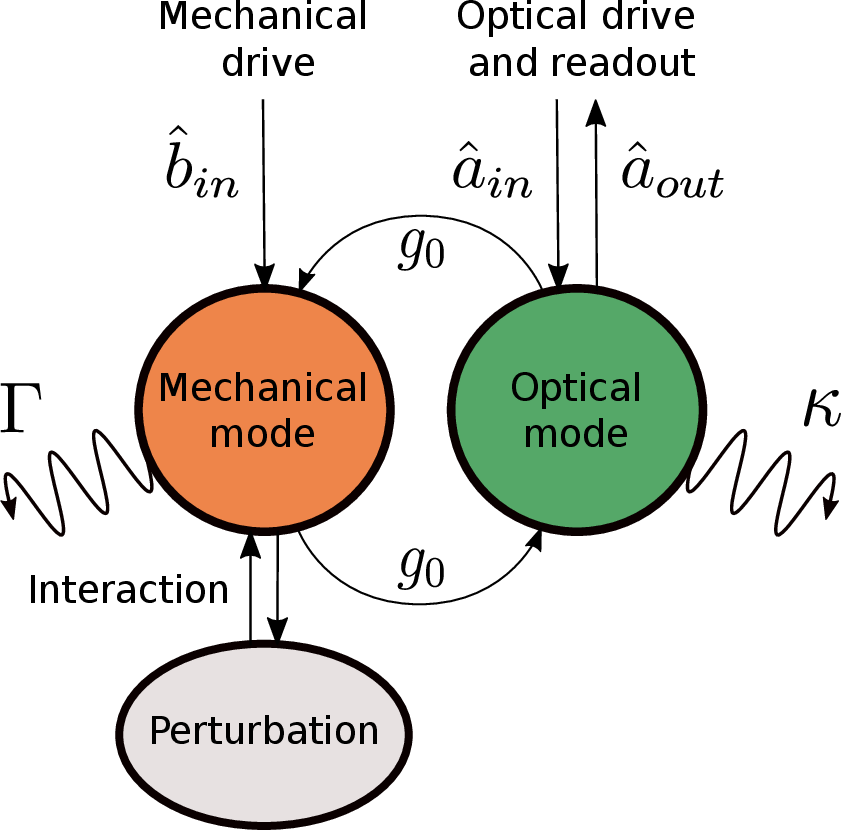}
\caption{Schematic of an optical mode with coupling $g_0$ to a mechanical mode perturbed by an external interaction, e.g. the tip-surface force in AFM. Both modes are driven and experience dissipation.
The mechanical mode via $\hat b_{\mathrm{in}}$ and $\Gamma$, and the optical mode via $\hat{a}_{in}$ and $\kappa$. Mechanical displacement is read out via the optical mode through $\hat a_{\mathrm{out}}$.}
\label{fig:schematic}
\end{figure}
Cavity optomechanical systems have proven to be a good platform for force sensing~\cite{aspelmeyer2014review, Liu2012}.
Their versatility in size, mechanical frequency and physical implementation has lead to their use in a wide range of applications, from gravitational wave detection~\cite{Abbott2016ligo}, to micrometer scale accelerometers~\cite{Krause2012} and tests of fundamental physics~\cite{Bose2023}.
While they have achieved remarkable sensitivity, there will always be uncertainties from fluctuations, i.e. noise.
Two competing sources of noise, imprecision noise (shot noise) and backaction noise, are minimized at the so-called standard quantum limit (SQL), where each noise source adds an equivalent of a quarter of a phonon of noise energy~\cite{Clerk2010}.

Backaction can be evaded with a quantum non-demolition measurement, where the measured observable is a constant of motion~\cite{Braginsky1980}.
Measurements that fulfil this criteria include momentum measurement of a free mass, or the phonon occupancy of an optomechanical system with quadratic coupling~\cite{Thompson2008}.
Another alternative is to measure only one of the two quadratures of the mechanical motion which are decoupled during free evolution~\cite{Thorne1978} while directing backaction to the other quadrature.
Such a backaction evading measurement has been achieved using a two-tone pump in both the microwave~\cite{hertzberg2009bae, Suh2014} and optical~\cite{Shomroni2019} regime.
In addition,  squeezing of mechanical motion below the zero point fluctuation has been demonstrated with a similar pump scheme~\cite{wollman2015}.
Other techniques exploit quantum correlations~\cite{Mason2019} and parametric amplification~\cite{motazedifard2019force} to improve sensitivity.

The possibility of reaching SQL make cavity optomechanical systems attractive for high-precision measurements and quantum sensing.
In particular, force sensing and force gradient sensing with optomechanical systems have been a major topic of interest~\cite{armata2017quantum, qvarfort2018gravimetry, rademacher2020quantum, Liu2012}.
Building on backaction evasion, phase-sensitive detection of mechanical motion has been demonstrated with a coherently driven mechanical mode~\cite{hertzberg2009bae, roos2023}.
A high-quality factor (high-Q) mechanical resonator that is coherently driven on resonance experiences a large change in phase for a small change in the force gradient~\cite{Smith1995, Albrecht1991}.
Sensing force gradients with a driven mechanical oscillation is in many cases superior to sensing force via quasi-static displacement of the test mass.
Theoretical proposals to specifically sense force gradients with levitated nano-particles exist~\cite{Rudolph2022, Li2022}, and realisations of force gradient sensors include optomechanical systems based on nanowires~\cite{Fogliano2021}.

One application of force gradient sensing is dynamic atomic force microscopy (AFM).
In the most standard setup a high-$Q$ cantilever with a sharp tip is actuated close to its resonance frequency and its motion is read-out by optical means.
Changes in motion result from changes in the local forces acting on the tip.
Scanning the tip over a surface, one can reconstruct interaction forces and surface topography at the atomic scale.
New proposals and designs for AFM have emerged which take advantage of the principles of cavity optomechanics~\cite{he2020optomechanical}, such as microdisk resonators coupled to waveguides~\cite{Srinivasan2011, Schwab2022}, nanomechanical beams~\cite{gavartin2012hybrid}, a doubly clamped cantilever~\cite{liu2012wide}, a Fabry-P\'erot cavity~\cite{melcher2014self}, membrane-based scanning~\cite{Hälg2021}, a microdisk coupled to a cantilever~\cite{doolin2014multidimensional}, and a cantilever coupled to a superconducting microwave readout~\cite{roos2023}.

In this work we investigate the theoretical sensitivity of force gradient detection with a backaction-evading measurement of a coherently driven mechanical mode.
The experiment is schematically depicted in Fig.~\ref{fig:schematic}.
By controlling the phase of a two-tone optical pump ($\hat a_{\mathrm{in}}$) relative to the mechanical drive ($\hat b_{\mathrm{in}}$) we encode the mechanical phase shift in the optical output ($\hat a_{\mathrm{out}}$) and select an optimal operating point for force gradient sensing.
While our work is motivated by improving dynamic AFM, the general measurement scheme and theoretical analysis applies to a variety of force gradient sensing applications.

The paper is structured as follows: In Sec.~\ref{sec:model} we describe the system and Hamiltonian. Section~\ref{sec:force:grad:scheme} describes the pumping scheme and the classical and quantum dynamics. In Sec.~\ref{sec:results} we present the classical response and noise analysis. Finally we summarize and conclude in Sec.~\ref{sec:conclusions}.

\section{Model}\label{sec:model}

Standard cavity optomechanics couples a high-frequency electromagnetic mode with resonance frequency $\omega_c$, to a mechanical mode with resonance frequency $\omega_m$.
We use the term 'optical' to refer to the high-frequency mode, which may be microwave or radio frequency.
The coupling results in the modulation of $\omega_c$ through mechanical displacement $x$, described by
\begin{align}\label{eq:lin:coupling}
    \omega_c(x) \approx \omega_c + \frac{\partial \omega_c}{\partial x}x + \ldots ,
\end{align}
where the higher order terms can be neglected if the modulation is small.
Promoting $x$ to an operator, the total Hamiltonian reads,
\begin{align}\label{eq:lin:ham}
	\hat H = \hbar \omega_c \hat a^\dag \hat a + \frac{\hat p^2}{2m_\mathrm{eff}}+\frac{1}{2}m_\mathrm{eff}\omega_m^2\hat x^2 - \frac{\hbar  g_0}{x_\mathrm{zpf}} \hat a^\dag \hat a \hat x,
\end{align}
where $m_\mathrm{eff}$ is the effective mass,  $x_\mathrm{zpf} = \sqrt{\hbar/(2m_\mathrm{eff}\omega_m)}$ the zero point fluctuation of the mechanical mode, and $g_0  = -x_\mathrm{zpf}\partial\omega_c/\partial x $ the single-photon optomechanical  coupling rate.
The operators $\hat a$ and $\hat a^\dag$ are the annihilation and creation operators of the optical field,  obeying the canonical commutator relation $[\hat a, \hat a^\dag] = 1$.
The mechanical position $\hat x = x_\mathrm{zpf}(\hat b+ b^\dag)$  and momentum $\hat p= -im_{\mathrm{eff}}\omega_mx_\mathrm{zpf}(\hat b-\hat b^\dag)$ operators obey $[\hat x, \hat p]  = i \hbar $.

We consider the example of non-contact atomic force microscopy where the mechanical resonator is a cantilever beam, clamped at one end and having a sharp tip at its free end.
The tip undergoes small oscillations $x(t)$ about its equilibrium position.
The AFM scanner controls the distance $h$ between this equilibrium position and the surface.
When the tip is close to the surface it experiences an interaction approximated by a van der Waals potential between a sphere and a flat surface,
\begin{equation}
    U(x) = -\frac{HR_\mathrm{tip}}{6(h+x)},
\end{equation}
where $H$ is the Hamaker constant and $R_\mathrm{tip}$ is the tip radius~\cite{giessibl2003review}.
We Taylor expand the potential and truncate after second order, valid for a small oscillation $x$. Again, promoting $x$ to an operator we find,
\begin{align}
    U(\hat x) = -\frac{H R_{\text{tip}}}{6 h} \left( 1 - \frac{\hat x}{h} + \frac{\hat x^2}{h^2} \right) + \mathcal{O}(\hat x^3).
\end{align}
The first term is simply a static shift which does not impact the equations of motion and can be disregarded.

The Hamiltonian with the truncated tip-surface force reads,
\begin{align}
	\hat H &= \hbar \omega_c \hat a^\dag \hat a + \frac{\hat p^2}{2m_\mathrm{eff}}+\frac{1}{2}m_\mathrm{eff}\omega_m^2\hat x^2 - \frac{\hbar  g_0}{x_\mathrm{zpf}} \hat a^\dag \hat a \hat x  \nonumber \\
	&\qquad - F_1\hat x - F_2\hat x^2,
\end{align}
where $F_1 = -HR_{\mathrm{tip}}/(6h^2)$ is a static tip-surface force which moves the equilibrium position of the mechanical mode.
The force gradient $F_2 = HR_\mathrm{tip}/(6h^3)$ results in an effective shift of the mechanical resonance frequency,
\begin{align}
\omega_{\mathrm{eff}} =  \sqrt{ \omega_m^2 -2F_2 /m_\mathrm{eff} }.
\end{align}
Rewriting the Hamiltonian with the effective resonance frequency,
\begin{align}
	\hat H &= \hbar \omega_c \hat a^\dag \hat a + \frac{\hat p^2}{2m_\mathrm{eff}}+\frac{1}{2}m_\mathrm{eff}\omega_{\mathrm{eff}}^2(F_2)\hat x^2 \nonumber \\
	&\qquad - \frac{\hbar  g_0}{x_\mathrm{zpf}} \hat a^\dag \hat a \hat x  - F_1\hat x,
\end{align}
we go to a frame rotating at an optical frequency which is detuned from cavity resonance  $\omega_p = \omega_c+\Delta$.
In terms of mechanical annihilation and creation operators
the Hamiltonian reads,
\begin{align}
\hat{H}_{rot} &= -\hbar\Delta\hat{a}^\dag\hat{a}+\hbar\omega_\mathrm{eff}(F_2)\hat{b}^\dag\hat{b} -\hbar g_0\hat{a}^\dag\hat{a}(\hat{b}+\hat{b}^\dag) \nonumber \\
&\qquad -F_1x_\mathrm{zpf}(\hat{b}+\hat{b}^\dag ).
\end{align}
Note that the coupling $g_0$ depends on the mechanical resonance frequency through $x_\mathrm{zpf}$, but here we treat it as a constant, considering only small changes of $\omega_m$.

\section{Force gradient sensing}\label{sec:force:grad:scheme}

\subsection{Optical pumps and mechanical drive}

We consider the case where the  mechanical mode is coherently driven at a frequency $\omega_d$ which is close to $\omega_\mathrm{eff}$.
The cavity is pumped with two tones $\omega_\pm=\omega_p \pm \omega_d$.
The resulting sidebands from the optomechanical interaction interfere at their center frequency $\omega_p$, where the amplitude of the response depends on the phase of the mechanical motion.
The phase of the mechanical oscillator is thus detected in relation to the phase of the beating envelope realized by the two optical pumps.
We allow for a slight detuning $\Delta$ of  $\omega_p$ from the cavity resonance: $\omega_p = \omega_c + \Delta$.
The optical input field reads
\begin{align}
\hat a_\mathrm{in} =  \bar{a}_\mathrm{in_-} e^{ -i \omega_- t } + \bar{a}_\mathrm{in_+}e^{- i \omega_+ t} +  \hat d_\mathrm{in},
\end{align}
where the first two terms are the classical pump signals, and $\hat{d}_\mathrm{in}$ accounts for any input fluctuations.
In a backaction evading measurement the pumps should be placed with no detunings such that $\omega_p=\omega_c$ and $\omega_d=\omega_\mathrm{eff}$ leading to $\omega_{\pm} = \omega_c\pm\omega_\mathrm{eff}$, as any detuning induces backaction into the measurement.

The mechanical mode is driven by a single tone close to its effective resonance frequency, $\omega_d = \omega_{\mathrm{eff}}+\delta$ where $\delta$ is a small detuning.
The mechanical drive reads
\begin{align}
\hat b_{\mathrm{in}} &= \bar{\beta}_\mathrm{in} e^{- i \omega_d t} + \hat c_\mathrm{in},
\end{align}
with $ \hat c_\mathrm{in}$ accounting for fluctuations.
Here all drive terms have complex amplitudes denoted with a bar, representing classical signals, for example $\bar{\beta}_\mathrm{in} = |\bar{\beta}_\mathrm{in}|e^{-i\phi_m}$.

\subsection{Quantum and classical dynamics}

In the frame rotating with the pump center frequency $\omega_p$, we obtain the following Langevin equations of motion for the optical and mechanical modes,
\begin{align}
\dot{\hat{a}}&= i\Delta\hat{a}+ig_0\hat{a}(\hat{b}+\hat{b}^\dagger)-\frac{\kappa}{2}\hat{a} \nonumber \\
&\qquad -\sqrt{\kappa}(\bar{a}_{in_-}e^{i\omega_dt}+\bar{a}_{in_+}e^{-i\omega_dt})-\sqrt{\kappa}\hat{d}_\mathrm{in}, \nonumber\\
\dot{\hat{b}}&= -i\omega_\mathrm{eff}(F_2)\hat{b}+ig_0\hat{a}^\dagger\hat{a}+\frac{iF_1x_\mathrm{zpf}}{\hbar}-\frac{\Gamma}{2}\hat{b} \nonumber \\
&\qquad -\sqrt{\Gamma}|\bar\beta_\mathrm{in}|e^{-i\phi_m}e^{-i\omega_dt}-\sqrt{\Gamma}\hat{c}_\mathrm{in}, \label{eq:full:langevin}
\end{align}
where $\kappa$ and $\Gamma$ are the total optical and mechanical dissipation rates, respectively.
We now assume that the internal cavity losses are small in comparison to the external decay rate, such that $\kappa_\mathrm{ext}\approx \kappa$.
In other words, the force sensor is designed for an over-coupled cavity in which damping is dominated by useful signal flowing to the next stage in the detection chain.

We model the optical and mechanical response as small quantum fluctuations around their respective classical mean,
\begin{align}\label{eq:lin:opt:mech}
    \hat a &= \bar{\alpha} + \hat d,  \nonumber \\
    \hat b &=  \bar{\beta} + \hat c ,
\end{align}
where the expectation values $\bar{\alpha} = \braket{\hat{a}}$ and $\bar{\beta} = \braket{\hat{b}}$ correspond to the classical oscillatory motion of the optical and mechanical modes respectively.
Inserting Eq.~\eqref{eq:lin:opt:mech} into Eq.~\eqref{eq:full:langevin}, we find the Langevin equations for the quantum fluctuations,
\begin{align}\label{eq:quantum:fluct}
\dot{\hat{d}}&= i\Delta\hat{d}+ig_0(\bar{\alpha}+\hat{d})(\bar{\beta}+\hat{c}+\bar{\beta}^*+\hat{c}^\dagger) -\frac{\kappa}{2}\hat{d}-\sqrt{\kappa}\hat{d}_\mathrm{in},  \nonumber\\
\dot{\hat{c}}&= -i\omega_\mathrm{eff}(F_2)\hat{c}+ig_0(\bar{\alpha}^*+\hat{d}^\dagger)(\bar{\alpha}+\hat{d})-\frac{\Gamma}{2}\hat{c}-\sqrt{\Gamma}\hat{c}_\mathrm{in},
\end{align}
and the classical motion
\begin{align} \label{eq:classical:diff:eqs}
\dot{\bar{\alpha}}&= i\Delta\bar{\alpha}+ig_0\bar{\alpha}(\bar{\beta}+\bar{\beta}^*)-\frac{\kappa}{2}\bar{\alpha} \nonumber \\
&\qquad -\sqrt{\kappa}(\bar{a}_{in_-}e^{i\omega_dt}+\bar{a}_{in_+}e^{-i\omega_dt}),  \nonumber \\
\dot{\bar{\beta}}&= -i\omega_\mathrm{eff}(F_2)\bar{\beta}+ig_0|\bar{\alpha}|^2+\frac{iF_1x_\mathrm{zpf}}{\hbar}-\frac{\Gamma}{2}\bar{\beta} \nonumber \\
&\qquad -\sqrt{\Gamma}|\bar{\beta}_\mathrm{in}|e^{-i\phi_m}e^{-i\omega_dt}.
\end{align}
In the usual two-tone backaction evading measurement~\cite{bowen2015quantumoptomechanics} these equations do not include the terms with $F_1$, $F_2$ or  $\bar{\beta}_\mathrm{in}$, and can therefore be solved analytically in the frequency domain.
In our case, the introduction of a perturbing force and a mechanical drive requires numerical treatment of Eq.~\eqref{eq:classical:diff:eqs}.

\section{Results}\label{sec:results}
Solving the classical equations of motions in Eq.~\eqref{eq:classical:diff:eqs}
we show how the force gradient $F_2$ is detected from the cavity field at $\omega_p$.
Using Eq.~\eqref{eq:quantum:fluct} we model the noise and compare it to the usual backaction evasion case.

\subsection{Modelling the classical response}

\begin{figure*}[ht]
\includegraphics[width=17.2cm]{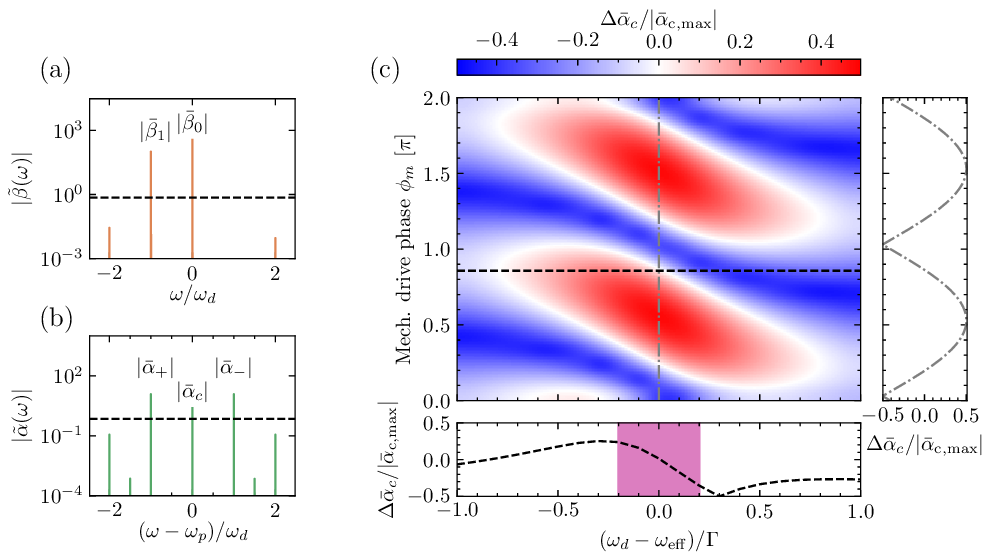}%
\caption{(a) and (b) Fourier transform of the solutions to the mechanical and optical classical equations of motion where the frequency components included in the following noise analysis are labeled. The black dashed lines indicate the square root of the zero point fluctuations. (c) Change in magnitude of the optical coherent tone  $|\bar{\alpha}_c|$ compared to the setpoint value $|\bar{\alpha}_\text{c,stp}|$ and normalized to the maximum value $|\bar{\alpha}_\mathrm{c,max}|$ as a function of mechanical drive phase $\phi_m$ and effective mechanical resonance frequency $\omega_{\mathrm{eff}}$.
To the right a vertical cut is shown detailing the interference pattern and on the bottom a horizontal cut where the purple region is highlighting the possible region of operation.
} %
\label{fig:fig2}
\end{figure*}

We solve the classical equations of motion (Eq.~\eqref{eq:classical:diff:eqs}) by numerical integration using the \texttt{scipy} function \texttt{ODEint}~\cite{scipy2020}.
We simulate a tip-surface interaction given in Section~\ref{sec:model} with $HR_\mathrm{tip}=3.55\cdot10^{-28} $ $\mathrm{Jm}$, for example achieved with $H= $ \SI{0.071}{\atto\joule} and $R_\mathrm{tip} = $ \SI{5.0}{\nano\meter}, with an optical pump amplitude of $\bar{a}_{in_\pm}\approx1.62\cdot10^5$ $\sqrt{\mathrm{photons}/\mathrm{s}}$, corresponding to the power needed to reach SQL for an optical (microwave) mode with resonance frequency $\omega_c/2\pi = $ \SI{4.5}{\giga\hertz} and total decay rate $\kappa/2\pi= $ \SI{1}{\mega\hertz}, coupled to a mechanical mode with resonance frequency $\omega_m/2\pi =$ \SI{5.37}{\mega\hertz}, total decay rate $\Gamma/2\pi= $ \SI{2.3}{\kilo\hertz}, effective mass $m_\mathrm{eff} =$ \SI{54}{\pico\gram}, with an optomechanical coupling rate $g_0/2\pi = $ \SI{1}{\kilo\hertz}.
The large mechanical linewidth is chosen to reduce computational time.
We apply a mechanical drive to yield an oscillation amplitude corresponding to 200 $x_\mathrm{zpf}$.
We set the unperturbed equilibrium position of the mechanical oscillator to $h = $ \SI{0.5}{\nano\meter}, approaching and retracting from the surface to obtain a mechanical frequency shift of one linewidth $\omega_\mathrm{d}\pm\Gamma$.

The Fourier transform of the solutions of the mechanical motion and the optical field are plotted in Fig.~\ref{fig:fig2}(a) and (b) for a mechanical drive phase $\phi_m = 0.86 \pi$ and $\omega_\mathrm{eff}=\omega_d$.
The approximated optical and mechanical fields reads
\begin{align}
\bar{\alpha}(t) &\approx \bar{\alpha}_{-}e^{i\omega_dt}+\bar{\alpha}_c+ \bar{\alpha}_{+}e^{-i\omega_dt}, \label{eq:opt:solution}\\
\bar{\beta}(t) &\approx \bar{\beta}_0+\bar{\beta}_1e^{-i\omega_dt}, \label{eq:mech:solution}
\end{align}
where higher-order frequency mixing products are neglected.
The term $\bar{\beta}_0$ is due to the static force which induces a shift in the optical resonance.
The term $\bar\beta_1$ is the response to the mechanical drive modulating the cavity resonance.
In the following analysis $2g_0|\bar\beta_1|\ll \kappa$ so that the linearization in Eq.~\eqref{eq:lin:coupling} is valid~\cite{Law1995}.

The quantity of interest is the term $|\bar{\alpha}_c|$ in Eq.~\eqref{eq:opt:solution} describing the coherent optical response at center frequency $\omega_p$, which is the result of the interference of the upper mechanical sideband from the red pump, and the lower mechanical sideband from the blue pump.
The magnitude of $|\bar{\alpha}_c|$ depends on the phase of the mechanical mode in relation to the phase of the beat created by the two optical pumps.
We change $|\bar{\alpha}_c|$ either through the mechanical drive phase $\phi_m$, or effective mechanical resonance frequency $\omega_\text{eff}(F_2)$ through the force gradient  by changing the distance to the surface $h$.

In Fig.~\ref{fig:fig2}(c) we plot the change in magnitude $\Delta\bar{\alpha}_c= |\bar{\alpha}_c|-|\bar{\alpha}_\mathrm{c,stp}|$ versus $\phi_m$ and $\omega_\mathrm{eff}$,
where $|\bar{\alpha}_\mathrm{c,stp}|$ is a setpoint for the AFM scanning feedback that controls $h$.
A cut along the vertical dash-dotted line reveals the interference between the two mechanical sidebands, shown in the rightmost panel.
The horizontal dashed line represents a typical setpoint for feedback, locking to the mechanical phase in the middle of an interference fringe.
We see in the lower panel that the change in response is monotonic for a small shift of mechanical resonance frequency $|\omega_d-\omega_\mathrm{eff}|\leq 0.2\Gamma$ (purple, shaded area).
The extended region of monotonic response to changes in $\omega_\mathrm{eff}$ makes $|\bar\alpha_\mathrm{c,stp}|$ a useful setpoint for scanning feedback.

\subsection{Noise analysis}

\begin{figure*}[ht]
\includegraphics[width=17.2cm]{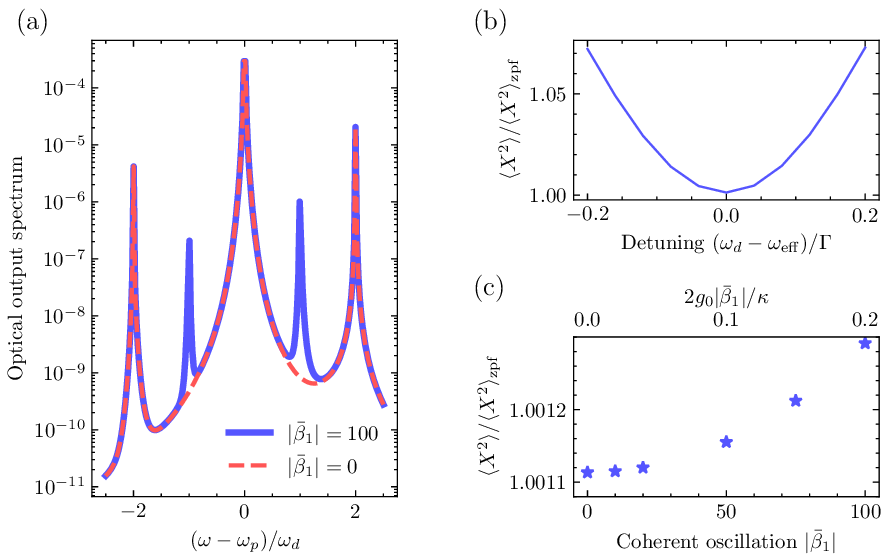}%
\caption{(a) Optical output spectrum detailing the difference between the pure backaction evasion  with no mechanical drive $|\bar\beta_1|=0$, and force gradient sensing with a mechanical drive $|\bar\beta_1|=100$. The addition of a coherent drive does not significantly add noise. (b) Quadrature fluctuations of the mechanical mode as a function of the detuning, calculated for $|\bar\beta_1|=100$. The range of detuning is chosen to match the monotonic regime of the classical response. (c) Quadrature fluctuations normalized to the zero point fluctuation plotted as a function of $|\bar\beta_1|$ for the case of resonant drive ($\omega_d=\omega_\mathrm{eff}$).}%
\label{fig:fig3}
\end{figure*}

In the usual backaction evading measurement the two-tone pumped optical cavity detects one of the mechanical quadratures without adding fluctuations to that quadrature.
However, in dynamic AFM where the mechanical mode is coherently driven, the combination of both optical pumping and mechanical driving leads to couplings which channel backaction fluctuations on to the detected quadrature.
When examining quantum limits to sensing force gradients, it is important to consider the impact of these additional noise contributions.

Returning to the equations of motions for the fluctuating terms, we insert Eqs.~\eqref{eq:opt:solution} and~\eqref{eq:mech:solution} into Eq.~\eqref{eq:quantum:fluct} and linearize to obtain
\begin{align}\label{eq:langevin:optics:time}
    \dot{\hat{d}}&= i\tilde{\Delta}\hat{d}+ig_0\lbrack(\bar{\alpha}_-e^{i\omega_dt}+\bar{\alpha}_c + \bar{\alpha}_+e^{-i\omega_dt})\hat{c}+(\bar{\alpha}_-e^{i\omega_dt}\nonumber\\
    &\quad+\bar{\alpha}_c +\bar{\alpha}_+e^{-i\omega_dt})\hat{c}^\dagger+\hat{d}(\bar{\beta}_1e^{-i\omega_dt}+\bar{\beta}^*_1e^{i\omega_dt})\rbrack\nonumber\\
    &\quad-\frac{\kappa}{2}\hat{d}-\sqrt{\kappa}\hat{d}_\mathrm{in},
\end{align}
\begin{align}\label{eq:langevin:mechanics:time}
    \dot{\hat{c}}&= -i\omega_\mathrm{eff}(F_2)\hat{c}+ig_0\lbrack(\bar{\alpha}_-^*e^{-i\omega_dt}+\bar{\alpha}^*_c + \bar{\alpha}^*_+e^{i\omega_dt})\hat{d}\nonumber\\
    &\quad+(\bar{\alpha}_-e^{i\omega_dt}+\bar{\alpha}_c + \bar{\alpha}_+e^{-i\omega_dt})\hat{d}^\dagger\rbrack-\frac{\Gamma}{2}\hat{c}-\sqrt{\Gamma}\hat{c}_\mathrm{in},
\end{align}
where $\tilde{\Delta} = \Delta+g_0(\bar{\beta}_0+\bar{\beta}_0^*)$.
At the setpoint of operation the placement of the pump center frequency can be adjusted such that $\tilde\Delta=0$.
As the AFM scanner moves and the static displacement $\bar\beta_0$ changes, a small non-zero detuning $\tilde\Delta$ needs to be taken into account.
We see that both the optical (Eq.~\eqref{eq:langevin:optics:time}) and mechanical (Eq.~\eqref{eq:langevin:mechanics:time}) fluctuations are periodic with $\omega_d$, complicating the analysis.
To solve for these fluctuations we turn to Floquet theory using the same approach as in~\cite{malz2016floquet, malz2016optomechanical} where we expand according to
\begin{align}
    \hat{d}(t) = \sum_{n=-\infty}^{\infty} e^{in\omega_dt}\hat{d}^{(n)}(t),\quad
    \hat{d}^{\dag}(t) = \sum_{n=-\infty}^{\infty} e^{in\omega_dt}\hat{d}^{(n)\dag}(t).
\end{align}
where $\hat{d}^{(n)}$ denotes a Fourier operator (in either time or frequency domain) rotating at frequency $n \omega_d$.
A similar expansion is made for the mechanical operators.
Rewriting Eqs.~\eqref{eq:langevin:optics:time} and~\eqref{eq:langevin:mechanics:time} in the form of stationary Langevin equations of motion and Fourier transforming to the frequency domain gives
\begin{align}\label{eq:langevin:floquet}
    -i\omega\hat{d}^{(n)}  &= (i \tilde{\Delta} - i  n \omega_d) \hat d^{(n)} + i g_0( \bar{\alpha}_-  \hat x^{(n-1)} + \bar{\alpha}_+  \hat x^{(n+1)}) \nonumber\\
    &\quad+ i g_0 \bar{\alpha}_c \hat x^{(n)}+ ig_0( \bar{\beta}_1 \hat d^{(n+1)} +  \bar{\beta}_1^*\hat d^{(n-1)})  \nonumber \\
    &\quad   - \frac{\kappa}{2}  \hat d^{(n)} - \sqrt{\kappa}   \hat d^{(n)}_\mathrm{in}, \\
    -i\omega\hat{c}^{(n)} &= -i\lbrack\omega_\mathrm{eff}(F_2) +n \omega_d\rbrack \hat c^{(n)} + i g_0  ( \bar{\alpha}_- \hat d^{(n-1)\dag}  \nonumber\\
    &\quad+\bar{\alpha}_+\hat d^{(n+1)\dag})+ i g_0 ( \bar{\alpha}_-^* \hat d^{(n+1)} + \bar{\alpha}_+^*   \hat d^{(n-1)})\nonumber\\
    &\quad+ i g_0 ( \bar{\alpha}_c ^* \hat d^{(n)} + \bar{\alpha}_c \hat d^{(n)\dag})- \frac{\Gamma}{2} \hat c^{(n)} - \sqrt{\Gamma} \hat c_\mathrm{in}^{(n)},
\end{align}
where $\hat{x}^{(n)}=\hat{c}^{(n)}+\hat{c}^{(n)\dag}$.
For the full calculation see Appendix~\ref{app:floquet:analysis}.

To understand how the mechanical drive and force gradient impacts fluctuations, we first calculate the fluctuations of the mechanical mode and later the spectrum of the output optical field.
We define the mechanical quadrature rotating at frequency $\omega_d$, coupled to the optical field in the back action evading measurement as~\cite{aspelmeyer2014review}
\begin{align}
    \hat{X}= \frac{1}{\sqrt{2}}\left(\hat{c}e^{i\omega_dt}+\hat{c}^\dag e^{-i\omega_dt}\right).
\end{align}
The variance of this quadrature is (see Appendix~\ref{app:mech:fluct})
\begin{align}
    &\braket{\hat X^2}=\int \frac{d\omega}{2\pi} S^{(0)}_{XX}(\omega) = \frac{1}{2}\int \frac{d\omega}{2\pi} \lbrack S^{(-2)}_{cc}(\omega+\omega_d) \nonumber\\
    &+S^{(0)}_{cc^\dag}(\omega+\omega_d)+S^{(2)}_{c^\dag c^\dag}(\omega-\omega_d)+S^{(0)}_{c^\dag c}(\omega-\omega_d)\rbrack,
\end{align}
where the components of the power spectrum are defined as
\begin{align}\label{eq:fourier:spectrum}
    S_{OO'}^{(m)}(\omega) = \sum_n\int\frac{d\omega'}{2\pi}\braket{\hat O^{(n)}(\omega+n\omega_d)\hat O'^{(m-n)}(\omega')}.
\end{align}
To calculate the optical spectrum we first write the equations for the terms at $n\pm1$
\begin{align}
    \hat{d}^{(n+1)} &= ig_0\chi_c\boldsymbol{(}\omega-\omega_d ( n+1 )\boldsymbol{)}(\bar{\alpha}_-\hat{x}^{(n)}+\bar{\alpha}_+\hat{x}^{(n+2)}\nonumber\\
    &\quad +\bar{\alpha}_c\hat{x}^{(n+1)}+\bar{\beta}_1\hat{d}^{(n+2)}+\bar{\beta}_1^*\hat{d}^{(n)}),  \nonumber\\
    \hat{d}^{(n-1)} &= ig_0\chi_c\boldsymbol{(}\omega-\omega_d( n-1) \boldsymbol{)}(\bar{\alpha}_-\hat{x}^{(n-2)}+\bar{\alpha}_+\hat{x}^{(n)}\nonumber\\
    &\quad+\bar{\alpha}_c\hat{x}^{(n-1)}+\bar{\beta}_1\hat{d}^{(n)}+\bar{\beta}_1^*\hat{d}^{(n-2)}) , \label{eq:dnpm1}
\end{align}
where we have introduced the optical susceptibility $\chi_c(\omega) = \lbrack\kappa/2-i(\omega+\tilde\Delta)\rbrack^{-1}$.
Discarding the terms at $n\pm2$, valid for fast periodic modulation $ \kappa, \Gamma \ll\omega_d$~\cite{malz2019thesis}, we use Eq.~\eqref{eq:dnpm1} to arrive at an expression for the Fourier operators of the optical mode
\begin{align}
	    \hat{d}^{(n)}& = \chi_c'(\omega)\lbrack ig_0(\bar{\alpha}_-'x^{(n-1)}+\bar{\alpha}_+'x^{(n+1)}+\bar{\alpha}_c'x^{(n)}) \nonumber\\
	    &\quad -\sqrt{\kappa}\hat d_{\text{in}}^{(n)}\rbrack,
\end{align}
where the modified optical susceptibility is
\begin{align}
    \chi_c'(\omega) &= \lbrack\chi_c(\omega-\omega_dn)^{-1}+g_0^2|\bar{\beta}_1|^2\chi_c\boldsymbol{(}\omega-\omega_d (n+1 ) \boldsymbol{)}\nonumber\\
    &\quad+g_0^2|\bar{\beta}_1|^2\chi_c\boldsymbol{(}\omega-\omega_d ( n-1 )\boldsymbol{)}\rbrack^{-1},
\end{align}
and where we have defined
\begin{align}
    &\bar{\alpha}_-' = \bar{\alpha}_-+ig_0\bar{\beta}_1^*\chi_c\boldsymbol{(}\omega-\omega_d (n-1) \boldsymbol{)}\bar{\alpha}_c,\\
    &\bar{\alpha}_+' = \bar{\alpha}_++ig_0\bar{\beta}_1\chi_c\boldsymbol{(}\omega-\omega_d (n+1) \boldsymbol{)}\bar{\alpha}_c,\\
    &\bar{\alpha}_c' = \bar{\alpha}_c+ig_0\bar{\beta}_1\chi_c\boldsymbol{(}\omega-\omega_d (n+1) \boldsymbol{)}\bar{\alpha}_-\nonumber\\
    &\qquad+ig_0\bar{\beta}_1^*\chi_c\boldsymbol{(}\omega-\omega_d (n-1) \boldsymbol{)}\bar{\alpha}_+.
\end{align}
The measured power spectral density is given by the $m=0$ component of Eq.~\eqref{eq:fourier:spectrum}~\cite{malz2016floquet} (see Appendix~\ref{app:opt:spectrum})
\begin{align}\label{eq:optical_spectrum}
    S_{d^\dag d}^{(0)}(\omega) &= |\chi_c'(\omega)|^2g_0^2\lbrack|\bar{\alpha}_-'|^2S_{xx}^{(0)}(\omega+\omega_d)\nonumber\\
    &+|\bar{\alpha}_+'|^2S_{xx}^{(0)}(\omega-\omega_d)+\bar{\alpha}_-^{*\prime}\bar{\alpha}_+'S_{xx}^{(-2)}(\omega+\omega_d)\nonumber\\
    &+\bar{\alpha}_+^{*\prime}\bar{\alpha}_-'S_{xx}^{(2)}(\omega-\omega_d)+|\bar{\alpha}_c'|^2S_{xx}^{(0)}(\omega)\nonumber\\
    &+\bar{\alpha}_-^{*\prime}\bar{\alpha}_c'S_{xx}^{(-1)}(\omega+\omega_d)+\bar{\alpha}_+^{*\prime}\bar{\alpha}_c'S_{xx}^{(1)}(\omega-\omega_d)\nonumber\\
    &+\bar{\alpha}_c^{*\prime}\bar{\alpha}_-'S_{xx}^{(1)}(\omega)+\bar{\alpha}_c^{*\prime}\bar{\alpha}_+'S_{xx}^{(-1)}(\omega)\rbrack .
\end{align}
If $\bar{\alpha}_-=\bar{\alpha}_+$ the first four terms in Eq.~\eqref{eq:optical_spectrum} constitute the standard backaction evasion spectrum, where we have not used the rotating wave approximation.

Figure~\ref{fig:fig3}(a) shows a comparison of the optical output power spectral density $S_{d_{out}^\dag d_{out}}^{(0)}(\omega)=\kappa S_{d^\dag d}^{(0)}(\omega)$ for two cases: a coherent mechanical state driven with $|\bar\beta_1|=100$ at $\omega_d = \omega_\mathrm{eff}$, corresponding to a oscillation amplitude of 200 $x_\mathrm{zpf}$, and the usual backaction evading measurement where the mechanics in not driven, $|\bar\beta_1|=0$.
In both cases we set the mechanical phonon occupancy $\bar{n}_m^{th} = 0$ and cavity photon occupancy $\bar{n}_c^{th} = 0$.
We see that the spectrum in the driven case shows additional noise peaks around the two optical pump frequencies, stemming from the mixing of the coherent optical response $\bar{\alpha}_c$ with the mechanical ground state fluctuations.
The peaks at $\pm 2\omega_\mathrm{eff}$ are due to counter-rotating terms which are usually neglected in the rotating wave approximation.

To further see the effect of the mechanical resonance frequency shift from the influence of the tip-surface force gradient, we calculate the variance of the measured mechanical quadrature.
The mechanical resonance frequency shift results in an effective detuning of the two optical pumps from the optimal pumping configuration of the backaction evading measurement ($\omega_\pm=\omega_c \pm \omega_\mathrm{eff}$).
The effect of this detuning on the quadrature fluctuations was previously investigated in~\cite{wollman2015thesis} but without a mechanical drive.
Figure~\ref{fig:fig3}(b) shows the increase in the variance of the mechanical quadrature due to backaction noise as a function of the effective mechanical resonance frequency.
For no detuning the quadrature fluctuation is very close to that of the unmeasured mechanical state.
As the tip-surface force gradient introduces a change in the effective resonance frequency, additional backaction noise is introduced into the quadrature, leading to an increase of the noise around the center frequency component $\bar{\alpha}_c$ in the output optical field.
The amount of backaction in the unmeasured quadrature depends on the pump power. As the pump power increases above that at SQL, $P > P_\text{SQL}$, the measured quadrature noise increases more rapidly than shown in \ref{fig:fig3}(b) where $P = P_\text{SQL}$.

Figure~\ref{fig:fig3}(c) shows the fluctuations at zero detuning for several coherent oscillation amplitudes $|\bar\beta_1|$ of the mechanical mode.
We see that the additional backaction noise increases with the amplitude of the mechanical motion.
However, the increase in quadrature fluctuations from the detuning is much larger than that from the mechanical drive.
The upper $x$ axis gives the corresponding values of $2g_0|\bar\beta_1|/\kappa\ll1$ where the linearized optomechanical interaction is valid.

\subsection{Considerations for AFM}

The force measurement of interest plays a decisive role in the choice of probe and sensing mechanism.
Recent work proposes sensing force gradient by comparing the force acting on two spatially separated mechanical oscillators~\cite{Li2022, Rudolph2022}.
Our scheme, which is more appropriate to AFM, requires only a single driven mechanical resonator that detects force gradient along the direction of mechanical motion.
The experimental conditions of AFM, such as the static tip-surface distance and amplitude of oscillation, together with various noise sources, determine the force gradient sensitivity as expressed in Ref.~\cite{Smith1995}.

To improve the force gradient sensitivity and successfully avoid backaction  with the proposed detection scheme, the device should operate in the resolved sideband regime ($\kappa \ll \Omega_m$) and close to the ground state of the mechanical mode.
The former is routinely achieved with superconducting microwave circuits coupled to micro-mechanical resonators, whereas the latter is more difficult.
Noise contribution from counter-rotating terms are filtered by the cavity in the resolved sideband regime, whereas these terms contribute significantly and limit backaction evasion in the unresolved sideband regime.
The force gradient sensitivity when the mechanical mode is far from the ground state will be limited by thermal fluctuations.

Even for a thermal state with large fluctuations the equations of motion [Eqns. \eqref{eq:langevin:optics:time} and \eqref{eq:langevin:mechanics:time}] are valid for arbitrarily small coherent mechanical oscillation.
Small amplitude mechanical oscillation, $x_{\mathrm{zpf}}|\bar{\beta}_1|$, degrades the force gradient sensitivity, but enables AFM operation closer to the surface where force gradients are larger and higher spatial resolution is possible as the AFM scans over the surface.
Our detection scheme with a driven mechanical mode differs from standard backaction evasion in that the validity of the linear optomechanical coupling is bounded by the product of the bare coupling and coherent mechanical oscillation amplitude such that $2g_0|\bar{\beta}_1|\ll \kappa$.
As with standard backaction evasion, linear cavities can compensate for low $g_0$ by increasing the optical pump strength to reach SQL.
Depending on the optomechanical coupling and the desired working distance to the surface, the mechanical drive amplitude and optical drive power should be chosen appropriately.

\section{Summary and Conclusions} \label{sec:conclusions}

We examined a scheme for force gradient sensing with a driven mechanical mode where motion is detected with the two-tone backaction evading measurement of cavity optomechanics.
The detection principle gives a monotonic response to changes in force gradient which cause frequency shifts on the order of 20\% of the mechanical linewidth.
A smaller mechanical linewidth improves the force gradient sensitivity, but limits the region of monotonic response for feedback control of atomic force microscopy (AFM).

Using Floquet theory to analyze the equations of motion of fluctuating terms, we showed that quadrature fluctuations of the mechanical mode, as detected in the optical output spectrum, are slightly larger than that achieved with the usual backaction evading measurement [see Figs.~\ref{fig:fig3}(b) and (c)].
The increase in fluctuations comes mainly from the change of the effective mechanical resonance due to the force gradient, resulting in a detuning of the optical pumps from the optimal placement for backaction evasion.
The contribution of the mechanical drive to the increase in fluctuations is much less than that of the change of the mechanical resonance frequency.

The noise of the mechanical quadrature as detected at the center frequency $\omega_p$ can further be decreased by adjusting the amplitudes of the two optical pumps such that $|\bar{a}_{in-}| > |\bar{a}_{in+}|$ to produce a squeezed mechanical state.
However, un-equal pump amplitudes changes the region of monotonic response in the classical interference pattern, as complete destructive and constructive interference is no longer achieved.
This imbalance of the interferometer degrades the response to changes in force gradient.

Thermal mechanical fluctuations can be reduced by backaction cooling.
However, this type of cooling does not improve force sensitivity.
An alternating sequence of backaction cooling and force gradient measurement may improve sensitivity.
Here careful analysis is needed to compare the mechanical relaxation time with the required measurement as well as the impact by measuring when the system is not in thermal equilibrium.
We leave these topics to future work.

\section*{Acknowledgements}

The authors would like to thank Clara C.~Wanjura as well as the Quantum-Limited Atomic Force microscopy (QAFM) team for fruitful discussions:  T.~Glatzel, M.~Zutter, E.~Thol\'en, D.~Forchheimer, I.~Ignat, M.~Kwon, and D.~Platz. E.K.A., E.S., A.K.R.,  are supported by an EIC Pathfinder (FET Open) grant 828966 -- QAFM and the Swedish SSF grant ITM17-0343. S.Q.~is funded in part by the Wallenberg Initiative on Networks and Quantum Information (WINQ) and in part by the Marie Skłodowska--Curie Action IF programme \textit{Nonlinear optomechanics for verification, utility, and sensing} (NOVUS) -- Grant-Number 101027183. Nordita is supported in part by NordForsk.

\section*{Data Availability}
The data that support the findings of this study are openly available in Zenodo at \url{https://doi.org/10.5281/zenodo.11175476}, reference number 11175476.

\newpage

\begin{appendix}
\newpage

\begin{widetext}

\section{Solving the noise equation of motions}\label{app:floquet:analysis}

We start with the equations of motion, Eq.~\eqref{eq:quantum:fluct} from the main text, where the classical solution is inserted and the optical mode is in a frame rotating with $\omega_p$ and the mechanical mode is in the lab frame.
\begin{align}\label{eq:eom:opt}
\dot{\hat{d}}&= i\Delta\hat{d}+ig_0(\bar{\alpha}_-e^{i\omega_dt}+\bar{\alpha}_c + \bar{\alpha}_+e^{-i\omega_dt}+\hat{d})(\bar{\beta}_0+\bar{\beta}_1e^{-i\omega_dt}+\hat{c}+\bar{\beta}_0+\bar{\beta}^*_1e^{i\omega_dt}+\hat{c}^\dagger)\nonumber\\
&\qquad -\frac{\kappa}{2}\hat{d}-\sqrt{\kappa}\hat{d}_\mathrm{in}, \\
\dot{\hat{c}}&= -i\omega_\mathrm{eff}(F_2)\hat{c}+ig_0(\bar{\alpha}_-^*e^{-i\omega_dt}+\bar{\alpha}^*_c + \bar{\alpha}^*_+e^{i\omega_dt}+\hat{d}^\dagger)(\bar{\alpha}_-e^{i\omega_dt}+\bar{\alpha}_c + \bar{\alpha}_+e^{-i\omega_dt}+\hat{d})\nonumber\\
&\qquad-\frac{\Gamma}{2}\hat{c}-\sqrt{\Gamma}\hat{c}_\mathrm{in}.\label{eq:eom:mech}
\end{align}
Linearizing these equations gives
\begin{align}
\dot{\hat{d}}&= i\tilde{\Delta}\hat{d}+ig_0(\bar{\alpha}_-e^{i\omega_dt}\hat{c}+\bar{\alpha}_c\hat{c} + \bar{\alpha}_+e^{-i\omega_dt}\hat{c}+\bar{\alpha}_-e^{i\omega_dt}\hat{c}^\dagger+\bar{\alpha}_c\hat{c}^\dagger + \bar{\alpha}_+e^{-i\omega_dt}\hat{c}^\dagger+\hat{d}\bar{\beta}_1e^{-i\omega_dt}\nonumber\\
&\qquad+\hat{d}\bar{\beta}^*_1e^{i\omega_dt})-\frac{\kappa}{2}\hat{d}-\sqrt{\kappa}\hat{d}_\mathrm{in}, \\
\dot{\hat{c}}&= -i\omega_\mathrm{eff}(F_2)\hat{c}+ig_0(\bar{\alpha}_-^*e^{-i\omega_dt}\hat{d}+\bar{\alpha}^*_c\hat{d} + \bar{\alpha}^*_+e^{i\omega_dt}\hat{d}+\bar{\alpha}_-e^{i\omega_dt}\hat{d}^\dagger+\bar{\alpha}_c\hat{d}^\dagger + \bar{\alpha}_+e^{-i\omega_dt}\hat{d}^\dagger)\nonumber\\
&\qquad-\frac{\Gamma}{2}\hat{c}-\sqrt{\Gamma}\hat{c}_\mathrm{in},
\end{align}
and the conjugate
\begin{align}
\dot{\hat{d}}^\dagger&= -i\tilde{\Delta}\hat{d}^\dagger-ig_0(\bar{\alpha}_-^*e^{-i\omega_dt}\hat{c}^\dagger+\bar{\alpha}_c^*\hat{c}^\dagger + \bar{\alpha}_+^*e^{i\omega_dt}\hat{c}^\dagger+\bar{\alpha}_-^*e^{-i\omega_dt}\hat{c}+\bar{\alpha}_c^*\hat{c} + \bar{\alpha}_+^*e^{i\omega_dt}\hat{c}+\hat{d}^\dagger\bar{\beta}_1^*e^{i\omega_dt}\nonumber\\
&\qquad+\hat{d}^\dagger\bar{\beta}_1e^{-i\omega_dt})-\frac{\kappa}{2}\hat{d}^\dagger-\sqrt{\kappa}\hat{d}^\dagger_\mathrm{in}, \\
\dot{\hat{c}}^\dagger&= i\omega_\mathrm{eff}(F_2)\hat{c}^\dagger-ig_0(\bar{\alpha}_-e^{i\omega_dt}\hat{d}^\dagger+\bar{\alpha}_c\hat{d}^\dagger + \bar{\alpha}_+e^{-i\omega_dt}\hat{d}^\dagger+\bar{\alpha}^*_-e^{-i\omega_dt}\hat{d}+\bar{\alpha}_c^*\hat{d} + \bar{\alpha}_+^*e^{i\omega_dt}\hat{d})\nonumber\\
&\qquad-\frac{\Gamma}{2}\hat{c}^\dagger-\sqrt{\Gamma}\hat{c}^\dagger_\mathrm{in},
\end{align}
where $\tilde{\Delta} = \Delta+g_0(\bar{\beta}_0+\bar{\beta}_0^*)$. We note that both in the optical equation of motion as well as in the mechanical one there are rotating terms. By going into a rotating frame of the mechanics some of the terms could become stationary. However, some time-dependence remains, which necessitates further analysis.
Following~\cite{malz2016floquet} the Floquet theory stipulates that we expand the operators in terms of Fourier components:
\begin{align}\label{eq:floq:fourier:opt}
    &\hat d(t) = \sum_{n = - \infty}^\infty e^{i n \omega_d t}  \hat d^{(n)}(t),  &&
    \hat d^\dag(t) = \sum_{n = - \infty}^\infty e^{i n \omega_d t} \hat d^{(n)\dag}(t),  \\
    &\hat c(t) = \sum_{n = - \infty}^\infty e^{i n \omega_d t}  \hat c^{(n)}(t), &&
    \hat c^\dag(t) = \sum_{n = - \infty}^\infty e^{i n \omega_d t} \hat c^{(n)\dag}(t)\label{eq:floq:fourier:mech}
\end{align}
We may then also define
\begin{align}
    \hat d^{(n)}(\omega) &= \int_{- \infty}^\infty \mathrm{d} t e^{i \omega t} \hat d^{(n)} (t), \\
    \hat d^{(n)\dag}(\omega) &= \int_{- \infty}^\infty \mathrm{d} t e^{i \omega t} \hat d^{(n)\dag}(t).
\end{align}
We note that this convention implies $[\hat d^{(n)}(\omega)]^\dag = \hat d^{(-n)\dag}(- \omega)$. Moving along, when differentiate one of the Fourier components with respect to time, we must invoke the chain law:
\begin{align}
    \dot{\hat{d}} = \frac{d}{dt} \sum_{n = - \infty}^\infty e^{i n \omega_d t}  \hat d^{(n)}(t) =(i n \omega_d) \sum_{n = -  \infty}^\infty e^{i n \omega_d t}  \hat d^{(n)}(t) + \sum_{n = - \infty}^\infty e^{i n \omega_d t}  \dot{\hat{d}}^{(n)}(t).
\end{align}
We now insert Eqs.~\eqref{eq:floq:fourier:opt} and ~\eqref{eq:floq:fourier:mech} into Eqs.~\eqref{eq:eom:opt} and~\eqref{eq:eom:mech} to find
\begin{align}
	    \sum_{n = - \infty}^\infty e^{i n \omega_d t}  \dot{\hat{d}}^{(n)}(t)  &= \sum_{n = - \infty}^\infty e^{i n \omega_d t}( i\tilde{\Delta} - i n \omega_d )  \hat d^{(n)}(t) \nonumber\\
	    &\qquad + i g_0 \left( \bar{\alpha}_- e^{i \omega_d t }\sum_{n = - \infty}^\infty e^{i n \omega_d t}  \hat c^{(n)}(t) + \bar{\alpha}_+ e^{-i \omega_d t }\sum_{n = - \infty}^\infty e^{i n \omega_d t} \hat c^{(n)\dag}(t) \right) \nonumber \\
	    &\qquad + i g_0 \left( \bar{\alpha}_- e^{i \omega_d t }\sum_{n = - \infty}^\infty e^{i n \omega_d t}  \hat c^{(n)\dag}(t) + \bar{\alpha}_+ e^{-i \omega_d t }\sum_{n = - \infty}^\infty e^{i n \omega_d t} \hat c^{(n)}(t) \right) \nonumber \\
	    &\qquad+ i g_0 \bar{\alpha}_c\left(  \sum_{n = - \infty}^\infty e^{i n \omega_d t}  \hat c^{(n)}(t)  + \sum_{n = - \infty}^\infty e^{i n \omega_d t} \hat c^{(n)\dag}(t) \right) \nonumber \\
	    &\qquad+ i g_0\left(\bar{\beta}_1 e^{- i \omega_dt} \sum_{n = - \infty}^\infty e^{i n \omega_d t}  \hat d^{(n)}
	    (t) +  \bar{\beta}_1^* e^{i \omega_d t} \sum_{n = - \infty}^\infty e^{i n \omega_d t}  \hat d^{(n)}
	    (t)\right)  \nonumber \\
	     &\qquad - \frac{\kappa}{2}  \sum_{n = - \infty}^\infty e^{i n \omega_d t}  \hat d^{(n)}(t) - \sqrt{\kappa}  \sum_{n = - \infty}^\infty e^{i n \omega_d t}  \hat d^{(n)}_\mathrm{in}(t),  \nonumber \\
	     \sum_{n = - \infty}^\infty e^{i n \omega_d t}  \dot{\hat{c}}^{(n)}(t)&=  \sum_{n = - \infty}^\infty e^{i n \omega_d t} (-i \omega_\mathrm{eff}(F_2) - i n \omega_d)  \hat c^{(n)}(t) \nonumber\\
	     &\qquad+ i g_0  \left( \bar{\alpha}_- e^{i \omega_d t }\sum_{n = - \infty}^\infty e^{i n \omega_d t}  \hat d^{(n)}(t)+ \bar{\alpha}_+ e^{-i \omega_d t } \sum_{n = - \infty}^\infty e^{i n \omega_d t} \hat d^{(n)\dag}(t)\right) \nonumber \\
	     &\qquad + i g_0  \left( \bar{\alpha}_-^* e^{-i \omega_d t }\sum_{n = - \infty}^\infty e^{i n \omega_d t}  \hat d^{(n)}(t)+ \bar{\alpha}_+^* e^{i \omega_d t } \sum_{n = - \infty}^\infty e^{i n \omega_d t} \hat d^{(n)\dag}(t)\right)\nonumber\\
	   &\qquad+ i g_0 \left( \bar{\alpha}_c ^* \sum_{n = - \infty}^\infty e^{i n \omega_d t}  \hat d^{(n)}(t) + \bar{\alpha}_c  \sum_{n = - \infty}^\infty e^{i n \omega_d t} \hat d^{(n)\dag}(t)  \right)   \nonumber \\
	   &\qquad- \frac{\Gamma}{2} \sum_{n = - \infty}^\infty e^{i n \omega_d t}  \hat c^{(n)}(t)- \sqrt{\Gamma}\sum_{n = - \infty}^\infty e^{i n \omega_d t}  \hat c_\mathrm{in}^{(n)}(t).
\end{align}
If we absorb the rotating terms into the sums and then shift the indices we find the following set of coupled differential equations:
\begin{align}
    \dot{\hat{d}}^{(n)}  &= (i \tilde \Delta - i  n \omega_d) \hat d^{(n)} + i g_0 \left( \bar{\alpha}_-  \hat c^{(n-1)} + \bar{\alpha}_+  \hat c^{(n+1)\dag} \right)  + i g_0 \left( \bar{\alpha}_-  \hat c^{(n-1)\dag} + \bar{\alpha}_+  \hat c^{(n+1)} \right)\nonumber \\
    &\qquad+ i g_0 \bar{\alpha}_c\left(  \hat c^{(n)}  +  \hat c^{(n)\dag} \right)+ ig_0\left( \bar{\beta}_1 \hat d^{(n+1)} +  \bar{\beta}_1^*\hat d^{(n-1)}\right)    - \frac{\kappa}{2}  \hat d^{(n)} - \sqrt{\kappa}   \hat d^{(n)}_\mathrm{in},\nonumber \\
    \dot{\hat{c}}^{(n)} &= (-i \omega_\mathrm{eff}(F_2) - i  n \omega_d) \hat c^{(n)} + i g_0  \left( \bar{\alpha}_- \hat d^{(n-1)\dag} + \bar{\alpha}_+   \hat d^{(n+1)\dag}\right)  + i g_0  \left( \bar{\alpha}_-^* \hat d^{(n+1)} + \bar{\alpha}_+^*   \hat d^{(n-1)}\right)\nonumber\\
    &\qquad+ i g_0 \left( \bar{\alpha}_c ^* \hat d^{(n)} + \bar{\alpha}_c \hat d^{(n)\dag}  \right)- \frac{\Gamma}{2} \hat c^{(n)} - \sqrt{\Gamma} \hat c_\mathrm{in}^{(n)}.
\end{align}
The equivalent conjugate equations are
\begin{align}
    \dot{\hat{d}}^{(n)\dag} &= \left(-i \tilde\Delta - i  n \omega_d \right) \hat d^{(n)\dag} - i g_0 \left( \bar{\alpha}_-^* \hat c^{(n+1)\dag} + \bar{\alpha}_+^* \hat c^{(n-1)} \right) - i g_0 \left( \bar{\alpha}_-^* \hat c^{(n+1)} + \bar{\alpha}_+^* \hat c^{(n-1)\dag} \right) \nonumber \\
    &\qquad - i g_0 \bar{\alpha}_c^* \left(  \hat c^{(n)\dag} +  \hat c^{(n)} \right) - i g_0\left( \bar{\beta}_1^*  \hat d^{(n-1) \dag }  + \bar{\beta}_1 \hat d^{(n+1)\dag} \right) - \frac{\kappa}{2}\hat d^{(n)\dag} - \sqrt{\kappa} \hat d_\mathrm{in}^{(n)\dag}, \nonumber \\
   \dot{\hat{c}}^{(n)\dag} &= \left(i \omega_\mathrm{eff}(F_2) - i  n \omega_d\right) \hat c^{(n)\dag} - i g_0  \left( \bar{\alpha}_-^*  \hat d^{(n+1)} + \bar{\alpha}_+^* \hat d^{(n-1)} \right) - i g_0  \left( \bar{\alpha}_-  \hat d^{(n-1)\dag} + \bar{\alpha}_+ \hat d^{(n+1)\dag} \right)  \nonumber \\
    &\qquad - i g_0 \left( \bar{\alpha}_c\hat d^{(n)\dag} + \bar{\alpha}_c^*  \hat d^{(n)}  \right) - \frac{\Gamma}{2}\hat{c}^{(n)\dag} -\sqrt{\Gamma}\hat c_\mathrm{in}^{(n)\dag}.
\end{align}
Fourier transforming and casting this into a matrix equation we get for $n = 0$
\small
\begin{align}
	    \begin{pmatrix}
	    -i(\omega+\omega_d) +A^{(0)} & A^{(-1)}& 0\\
	    A^{(1)}&-i\omega+A^{(0)} &A^{(-1)}\\
	    0& A^{(1)} &-i(\omega-\omega_d) +A^{(0)}
	    \end{pmatrix}
	    \cross
	    \begin{pmatrix}
	    \hat{\mathbf{x}}^{(-1)}\\
	    \hat{\mathbf{x}}^{(0)}\\
	    \hat{\mathbf{x}}^{(1)}\\
	    \end{pmatrix}
	    =
	    \begin{pmatrix}
	    0\\
	    \hat{\mathbf{x}}_\mathrm{in}\\
	    0\\
	    \end{pmatrix},
\end{align}
\normalsize
where
\begin{align}
    A^{(0)} = \begin{pmatrix} -i\tilde{\Delta} + \frac{\kappa}{2}  & -ig_0 \bar{\alpha}_c& 0 &     -ig_0 \bar{\alpha}_c \\
    -ig_0 \bar{\alpha}_c^*& i \omega_\mathrm{eff} + \frac{\Gamma}{2} &  -ig_0 \bar{\alpha}_c & 0 \\
     0 & ig_0 \bar{\alpha}_c^* &  i \tilde{\Delta}+ \frac{\kappa}{2}  & ig_0 \bar{\alpha}_c^*\\
    ig_0 \bar{\alpha}_c^* & 0 & ig_0 \bar{\alpha}_c  & - i \omega_\mathrm{eff} + \frac{\Gamma}{2}
    \end{pmatrix},
\end{align}
and
\begin{align}
    A^{(-1)} &= -ig_0\begin{pmatrix}
    \bar{\beta}_1 & \bar{\alpha}_+ & 0 & \bar{\alpha}_+ \\
    \bar{\alpha}_-^* & 0 & \bar{\alpha}_+ & 0 \\
    0&  -\bar{\alpha}_-^* & -\bar{\beta}_1  & -\bar{\alpha}_-^* \\
    -\bar{\alpha}_-^* & 0 &-\bar{\alpha}_+ & 0
    \end{pmatrix}, \\
    A^{(1)} &= -ig_0\begin{pmatrix}
    \bar{\beta}_1^*  & \bar{\alpha}_- & 0 & \bar{\alpha}_-\\
    \bar{\alpha}_+^* & 0 &  \bar{\alpha}_- & 0 \\
    0 & -\bar{\alpha}_+^* & -\bar{\beta}_1^* &-\bar{\alpha}_+^*\\
    -\bar{\alpha}_+^*& 0 & -\bar{\alpha}_- & 0
    \end{pmatrix},
\end{align}
and
\begin{equation}
	    \hat{\mathbf{x}}^{(n)}= \begin{pmatrix}
		\hat{d}^{(n)} &\hat{c}^{(n)} &\hat{d}^{(n)\dag} & \hat{c}^{(n)\dag}
	    \end{pmatrix}^T,
\end{equation}
and
\begin{equation}
	    \hat{\mathbf{x}}_ \mathrm{in}= \begin{pmatrix}
		-\sqrt{\kappa}\hat{d}_\mathrm{in} &-\sqrt{\Gamma}\hat{c}_\mathrm{in} &-\sqrt{\kappa}\hat{d}_\mathrm{in}^\dag &-\sqrt{\Gamma} \hat{c}_\mathrm{in}^\dag
	    \end{pmatrix}^T,
\end{equation}
where we truncated at $n\pm2$.

\section{Mechanical fluctuations}\label{app:mech:fluct}

We define the mechanical quadrature as $\hat X = \frac{1}{\sqrt{2}}\left( \hat c e^{i\omega_dt+i\theta}+\hat c^\dag e^{-i\omega_dt-i\theta}\right)$.
The spectrum for the mechanical quadrature is
\begin{align}
    S_{XX}(\omega,t)&=\frac{1}{2}\int_{-\infty}^{\infty} d\tau e^{i\omega\tau}\langle(\hat c(t+\tau) e^{i\omega_d(t+\tau)+i\theta}\nonumber\\
    &\quad+\hat c^\dag(t+\tau) e^{-i\omega_d(t+\tau)-i\theta})(\hat c(t) e^{i\omega_dt+i\theta}+\hat c^\dag(t) e^{-i\omega_dt-i\theta})\rangle\nonumber\\
    &=\frac{1}{2}\sum_{n,m}\int_{-\infty}^{\infty} d\tau e^{i\omega\tau}\braket{\hat c^{(n)}(t+\tau)e^{in\omega_d(t+\tau)} e^{i\omega_d(t+\tau)+i\theta}\hat c^{(m)}(t)e^{im\omega_dt}e^{i\omega_dt+i\theta}}\nonumber\\
    &\qquad+\frac{1}{2}\sum_{n,m}\int_{-\infty}^{\infty} d\tau e^{i\omega\tau}\braket{\hat c^{(n)}(t+\tau)e^{in\omega_d(t+\tau)} e^{i\omega_d(t+\tau)+i\theta}\hat c^{(m)\dag}(t)e^{im\omega_dt}e^{-i\omega_dt-i\theta}}\nonumber\\
    &\qquad+\frac{1}{2}\sum_{n,m}\int_{-\infty}^{\infty} d\tau e^{i\omega\tau}\braket{\hat c^{(n)\dag}(t+\tau)e^{in\omega_d(t+\tau)} e^{-i\omega_d(t+\tau)-i\theta}\hat c^{(m)}(t)e^{im\omega_dt}e^{i\omega_dt+i\theta}}\nonumber\\
    &\qquad+\frac{1}{2}\sum_{n,m}\int_{-\infty}^{\infty} d\tau e^{i\omega\tau}\braket{\hat c^{(n)\dag}(t+\tau)e^{in\omega_d(t+\tau)} e^{-i\omega_d(t+\tau)-i\theta}\hat c^{(m)\dag}(t)e^{im\omega_dt}e^{-i\omega_dt-i\theta}}.
\end{align}
For each term individually, we have
\begin{align}
    &\sum_{n,m}\int_{-\infty}^{\infty} d\tau e^{i\omega\tau}\braket{\hat c^{(n)}(t+\tau)e^{in\omega_d(t+\tau)} e^{i\omega_d(t+\tau)+i\theta}\hat c^{(m)}(t)e^{im\omega_dt}e^{i\omega_dt+i\theta}}\nonumber\\
    &= \sum_{n,m} e^{i(n+m)\omega_dt}e^{2i\omega_dt}\int_{-\infty}^{\infty} d\tau e^{i\omega\tau}e^{i\omega_d\tau}e^{in\omega_d\tau}\braket{\hat c^{(n)}(t+\tau)\hat c^{(m)}(t)}e^{2i\theta},
\end{align}
and
\begin{align}
    &\sum_{n,m}\int_{-\infty}^{\infty} d\tau e^{i\omega\tau}\braket{\hat c^{(n)}(t+\tau)e^{in\omega_d(t+\tau)} e^{i\omega_d(t+\tau)+i\theta}\hat c^{(m)\dag}(t)e^{im\omega_dt}e^{-i\omega_dt-i\theta}}\nonumber\\
    &= \sum_{n,m} e^{i(n+m)\omega_dt}\int_{-\infty}^{\infty} d\tau e^{i\omega\tau}e^{i\omega_d\tau}e^{in\omega_d\tau}\braket{\hat c^{(n)}(t+\tau)\hat c^{(m)\dag}(t)},
\end{align}
and
\begin{align}
    &\sum_{n,m}\int_{-\infty}^{\infty} d\tau e^{i\omega\tau}\braket{\hat c^{(n)\dag}(t+\tau)e^{in\omega_d(t+\tau)} e^{-i\omega_d(t+\tau)-i\theta}\hat c^{(m)}(t)e^{im\omega_dt}e^{i\omega_dt+i\theta}}\nonumber\\
    &= \sum_{n,m} e^{i(n+m)\omega_dt}\int_{-\infty}^{\infty} d\tau e^{i\omega\tau}e^{-i\omega_d\tau}e^{in\omega_d\tau}\braket{\hat c^{(n)\dag}(t+\tau)\hat c^{(m)}(t)},
\end{align}
as well as
\begin{align}
    &\sum_{n,m}\int_{-\infty}^{\infty} d\tau e^{i\omega\tau}\braket{\hat c^{(n)\dag}(t+\tau)e^{in\omega_d(t+\tau)} e^{-i\omega_d(t+\tau)-i\theta}\hat c^{(m)\dag}(t)e^{im\omega_dt}e^{-i\omega_dt-i\theta}}\nonumber\\
    &= \sum_{n,m} e^{i(n+m)\omega_dt}e^{-2i\omega_dt}\int_{-\infty}^{\infty} d\tau e^{i\omega\tau}e^{-i\omega_d\tau}e^{in\omega_d\tau}\braket{\hat c^{(n)\dag}(t+\tau)\hat c^{(m)\dag}(t)}e^{-2i\theta}.
\end{align}
If we for each term let $m=m'-n$ and enforce that the power spectral density of the quadrature fluctuation is given by the time-independent contribution we find,
\begin{align}
    S_{XX}^{(0)}(\omega) = \frac{1}{2}\boldsymbol{(}S_{cc}^{(-2)}(\omega+\omega_d)e^{2i\theta}+ S_{cc^\dag}^{(0)}(\omega+\omega_d)+S_{c^\dag c}^{(0)}(\omega-\omega_d)+S_{c^\dag c^\dag}^{(2)}(\omega-\omega_d)e^{-2i\theta}\boldsymbol{)}
\end{align}

where the components are defined according to Eq.~\eqref{eq:fourier:spectrum} in the main text.

\section{Optical spectrum}\label{app:opt:spectrum}

The optical equation of motion is
\begin{align}
    \chi_c'^{-1}(\omega)\hat{d}^{(n)} = ig_0(\bar{\alpha}_-'\hat x^{(n-1)}+\bar{\alpha}_+'\hat x^{(n+1)}+\bar{\alpha}_c'\hat x^{(n)}) -\sqrt{\kappa}\hat d_{\text{in}}^{(n)},
\end{align}
and for the conjugate
\begin{align}
    (\chi_c'^{-1})^*(\omega)\hat{d}^{(n)\dag} = -ig_0(\bar{\alpha}_-'^*\hat x^{(n+1)}+\bar{\alpha}_+'^*\hat x^{(n-1)}+\bar{\alpha}_c'^*\hat x^{(n)}) -\sqrt{\kappa}\hat d_{\text{in}}^{(n)\dag}.
\end{align}
The optical spectrum is calculated by considering each of the mechanical contributions. For  $\hat x_a(t)=\hat x(t) e^{ia\omega_dt}=\sum_n \hat x^{(n-a)}(t)$ and $\hat x_b= \hat x(t) e^{i b\omega_dt}=\sum_n \hat x^{(n-b)}(t)$ we have
\begin{align}
    S_{x_a x_b}(\omega,t)&=\int_{-\infty}^{\infty} d\tau e^{i\omega\tau}\braket{\hat x(t+\tau) e^{ia\omega_d(t+\tau)}\hat x(t)e^{ib\omega_dt}}\nonumber\\
    &=\sum_{n,m}\int_{-\infty}^{\infty} d\tau e^{i\omega\tau}\braket{\hat x^{(n)}(t+\tau)e^{in\omega_d(t+\tau)} e^{ia\omega_d(t+\tau)}\hat x^{(m)}(t)e^{im\omega_dt}e^{ib\omega_dt}}\nonumber\\
    &=\sum_{n,m} e^{i(n+m)\omega_dt}e^{i(a+b)\omega_dt}\int_{-\infty}^{\infty} d\tau e^{i\omega\tau}e^{ia\omega_d\tau}e^{in\omega_d\tau}\braket{\hat x^{(n)}(t+\tau)\hat x^{(m)}(t)}\nonumber\\
    &=\sum_m S_{x_ax_b}^{(m)}(\omega)e^{im\omega_dt}.
\end{align}
If we let $m=m'-n$ we get
\begin{align}
    S_{x_a x_b}(\omega,t)&=\sum_{n,m'} e^{im'\omega_dt}e^{i(a+b)\omega_dt}\int_{-\infty}^{\infty} d\tau e^{i\omega\tau}e^{ia\omega_d\tau}e^{in\omega_d\tau}\braket{\hat x^{(n)}(t+\tau)\hat x^{(m'-n)}(t)}.
\end{align}
The last integral we can instead write as
\begin{align}
    &\sum_n\int_{-\infty}^{\infty} d\tau e^{i\omega\tau}e^{ia\omega_d\tau}e^{in\omega_d\tau}\braket{\hat x^{(n)}(t+\tau)\hat x^{(m'-n)}(t)} = \sum_n \int_{-\infty}^{\infty}\frac{d\omega'}{2\pi}\braket{\hat x^{(n)}(\omega+a\omega_d+n\omega_d)\hat x^{(m'-n)}(\omega')} \nonumber\\
    &\quad= S_{xx}^{(m')}(\omega+a\omega_d).
\end{align}
We therefore have
\begin{align}
    S_{x_a x_b}(\omega,t)&=\sum_m  S^{(m)}_{x_a x_b}(\omega) e^{im\omega_dt} = \sum_{m'} e^{im'\omega_dt}e^{i(a+b)\omega_dt}S_{xx}^{(m')}(\omega+a\omega_d).
\end{align}

For the steady state spectrum we require that $m'=-a-b $. This means that the contribution to the optical spectrum becomes
\begin{align}
    S_{x_a x_b}^{(0)}(\omega) = S_{xx}^{(-a-b)}(\omega+a\omega_d).
\end{align}
The spectrum of the optical mode thus becomes
\begin{align}
    S^{(0)}_{d^\dag d}(\omega) &= g_0^2|\chi_c'(\omega)|^2(|\bar{\alpha}_-'|^2S_{xx}^{(0)}(\omega+\omega_d)+|\bar{\alpha}_+'|^2S_{xx}^{(0)}(\omega-\omega_d)+\bar{\alpha}_-'^*\bar{\alpha}_+'S_{xx}^{(-2)}(\omega+\omega_d)\nonumber\\
    &\quad+\bar{\alpha}_+'^*\bar{\alpha}_-'S_{xx}^{(2)}(\omega-\omega_d)+|\bar{\alpha}_c'|^2S_{xx}^{(0)}(\omega)+\bar{\alpha}_-'^*\bar{\alpha}_c'S_{xx}^{(-1)}(\omega+\omega_d)+\bar{\alpha}_+'^*\bar{\alpha}_c'S_{xx}^{(1)}(\omega-\omega_d)\nonumber\\
    &\quad+\bar{\alpha}_c'^*\bar{\alpha}_-'S_{xx}^{(1)}(\omega)+\bar{\alpha}_c'^*\bar{\alpha}_+'S_{xx}^{(-1)}(\omega),
\end{align}
where
\begin{align}
    S_{xx}^{(m)}(\omega) &= \sum_n \int \frac{d\omega'}{2\pi}\braket{\hat x^{(n)}(\omega+n\omega_d)\hat x^{(m-n)}(\omega')}\nonumber\\
    &=\sum_n \int \frac{d\omega'}{2\pi}\braket{(\hat c^{(n)}(\omega+n\omega_d)+\hat c^{(n)\dag}(\omega+n\omega_d))(\hat c^{(m-n)}(\omega')+\hat c^{(m-n)\dag})(\omega')}.
\end{align}

\end{widetext}

\end{appendix}

\bibliographystyle{apsrev4-2}

\end{document}